\documentclass[aps,pre,preprint,showpacs]{revtex4-1}
\usepackage{booktabs}
\usepackage{graphicx}
\usepackage{sidecap}
\usepackage{dcolumn}
\usepackage{bm}
\usepackage{multirow} 
\usepackage{epstopdf} 
\usepackage{amsfonts}
\usepackage{amsmath}
\usepackage{setspace}
\usepackage[titletoc,toc]{appendix}
\usepackage{color}
\usepackage{hyperref}
\hypersetup{colorlinks=true,linkcolor=blue,citecolor=blue,breaklinks={true}}
\usepackage[draft]{changes} 

\font\bbfnt=msbm10

\def\bbR{\mbox{\bbfnt R}}

\newcommand{\mb}[1]{\mbox{\bfseries \itshape #1}}

\begin{document}


\title{Mixed measurements and the detection of phase synchronization in networks}

\author{Leonardo L. Portes}
\email{ll.portes@gmail.com}
\affiliation{Departamento de F\'isica,
Universidade Federal de Minas Gerais}
\affiliation{Programa de P\'os-Gradua\c{c}\~ao em Engenharia El\'etrica da
Universidade Federal de Minas Gerais --  
Av. Ant\^onio Carlos 6627, 31.270-901 Belo Horizonte MG, Brazil}

\author{Luis A. Aguirre} 
\affiliation{Departamento de Engenharia Eletr\^onica,
Universidade Federal de Minas Gerais}
\affiliation{Programa de P\'os-Gradua\c{c}\~ao em Engenharia El\'etrica da
Universidade Federal de Minas Gerais --  
Av. Ant\^onio Carlos 6627, 31.270-901 Belo Horizonte MG, Brazil}

\date{\today}

\begin{abstract}
Multivariate singular spectrum analysis (M-SSA), with a varimax rotation of eigenvectors, was recently proposed to provide detailed information about phase synchronization in networks of nonlinear oscillators without any a priori need for phase estimation. The discriminatory power of M-SSA is often enhanced by using only the time series of the variable that provides the best observability of the node dynamics. In practice, however, diverse factors could prevent one to have access to this variable in some nodes and other variables should be used, resulting in a mixed set of variables. In the present work, the impact of this mixed measurement approach on the M-SSA is numerically investigated in networks of R\"ossler systems and cord oscillators. The results are threefold. First, a node measured by a poor variable, in terms of observability, becomes virtually invisible to the technique. Second, a side effect of using a poor variable is that the characterization of phase synchronization clustering of the {\it other}\, nodes is hindered by a small amount. This suggests that, given a network, synchronization analysis with M-SSA could be more reliable by not measuring those nodes that are accessible only through poor variables. Third, global phase synchronization could be detected even using only poor variables, given enough of them are measured. These insights 
could be useful in defining measurement strategies for both experimental design and real world applications for use with M-SSA.
\end{abstract}

\pacs{05.45.-a,02.50.-r}

\maketitle

\section{Introduction}
\label{sec.introduction}

A long standing problem in the detection of phase synchronization (PS) is the need for a phase estimate or definition \cite{Rosenblum1996,Rosenblum1997,Pikovsky2001,Boccaletti2002,Osipov2007}. Given an oscillator with spiral dynamics, geometry-based definitions easily apply. Estimates that use Poincar\'e sections or projections onto specific phase-related planes can be sometimes applied to more complex dynamics (e.g., R\"ossler oscillator in the chaotic funnel regime and the Lorenz system, respectively). 

The multivariate singular spectrum analysis, along with a structured varimax rotation (svM-SSA), is a new and powerful technique for the analysis of synchronization phenomena in networks of coupled nonlinear oscillators \cite{Groth2011}. Such analysis provides detailed information about PS clustering by considering the ``skeleton" of the intrinsic oscillatory modes in data, and hence without any need for a phase estimate or definition, and has undergone intense development in recent years \cite{Pukenas2014,Groth2015,Portes2016b, Portes2016a}, and with applications to real-world data \cite{Feliks2013,Groth2017}.  

As originally proposed in \cite{Groth2011} the svM-SSA requires that, given a set of coupled oscillators, every state variable of each
oscillator be used in the analysis. This could turn out to be a limiting requirement in many practical
problems. An important way of reducing this constraint to some extent is to consider measuring only one -- but
always the same -- variable from each node. The question of which variable to use has been
investigated in \cite{Portes2016b}, where
it has been argued that the performance of svM-SSA in the characterization of PS is sually improved by 
using a single variable per node (e.g. a univariable approach) and choosing the variable that provides the best observability of the system dynamics \cite{Letellier2005}. Such a procedure was recently used
in the investigation of PS in networks of bursting neuron models \cite{Aguirre2017}. 

Although the flexibility attained by only having to measure one variable per node is a welcome feature,
it might still pose limitations in practice because 
it might not always be possible to measure the same variable from every oscillator in the network. 
Hence, the motivation of this work is to go a step further in relaxing the data requirements related
to the use of svM-SSA. In particular, the aim is to investigate the effect of using a {\it mixed}\, set of
variables in the analysis. This means that, for some oscillators, variables that convey less dynamical information will be used. The number and position in the network of such oscillators will be taken into account in the analysis. Is it still possible
to detect PS of the network as the number of poorly observed oscillators increases?

In fact, the results show that the behavior of oscillators (nodes) for which only a poor -- in terms of observability -- variable is measured is not detected by the svM-SSA, thus hindering the characterization of phase synchronization (PS) clustering. This effect is more evident in scenarios with more complex intrinsic node dynamics. Not only that, the characterization of PS clustering of nodes for which good observability variables are recorded is also hindered by a small amount. Hence the use of variables that convey
poor observability seems to be a problem for svM-SSA, at least in the investigated cases.

The paper is organized as follows. Section \ref{sec.method} provides background material related to the study rationale, the svM-SSA, numerical models and experimental design. Synchronization in networks of R\"ossler oscillators (spiral and funnel chaotic regimes) and Cord oscillators is analyzed through mixed measurements in Sec. \ref{sec.resulsts}. Final remarks and prescriptions are presented in Sec. \ref{sec.conclusion}.


\section{Method}
\label{sec.method}

\subsection{Statement of the problem}
\label{sec.problem}

Given a network of coupled oscillators $\dot{\mb{x}}_j=f_j(\mb{x}_j,\,\mb{v}),~j=1,..., J$, where
$\mb{x}\in\bbR^n$, $\mb{v}$ is a vector of coupling variables from neighbor oscillators.
It is assumed that there is a measuring function 
$h_j:\bbR^n\rightarrow \bbR$ for each oscillator, such that $h_j(\mb{x}_j)=s_j(t)$.
From the set of time series $s_j(t),~j=1,..., J$, it is desired to detect the formation of
phase-synchronized clusters in the network. 

In this work the following assumptions are made:

\begin{itemize}
\item[(a)] the oscillators are 3-dimensional ($n=3$) and of the same type with parameter mismatch;
\item[(b)] the measuring functions $h_j$ are {\it not}\, all the same;
\item[(c)] the functions $h_j$ will return one of the state variables of the $j$th oscillator or,
	eventually, none.
\end{itemize}

Hence the challenge is to assess PS of the network from a {\it mixed set of time series}.
This scenario -- schematically shown in Fig.~\ref{fig.scheme}(a) for a chain 
network -- represents a step forward with respect to the case in which the
{\it same}\, variable is measured from each oscillator  \cite{Portes2016b}.  
The numerical setup for investigating this problem is described next.

\subsection{Numerical experimental design}
\label{sec.design}


\begin{figure}[!tb]
\includegraphics[width=7.5 cm]{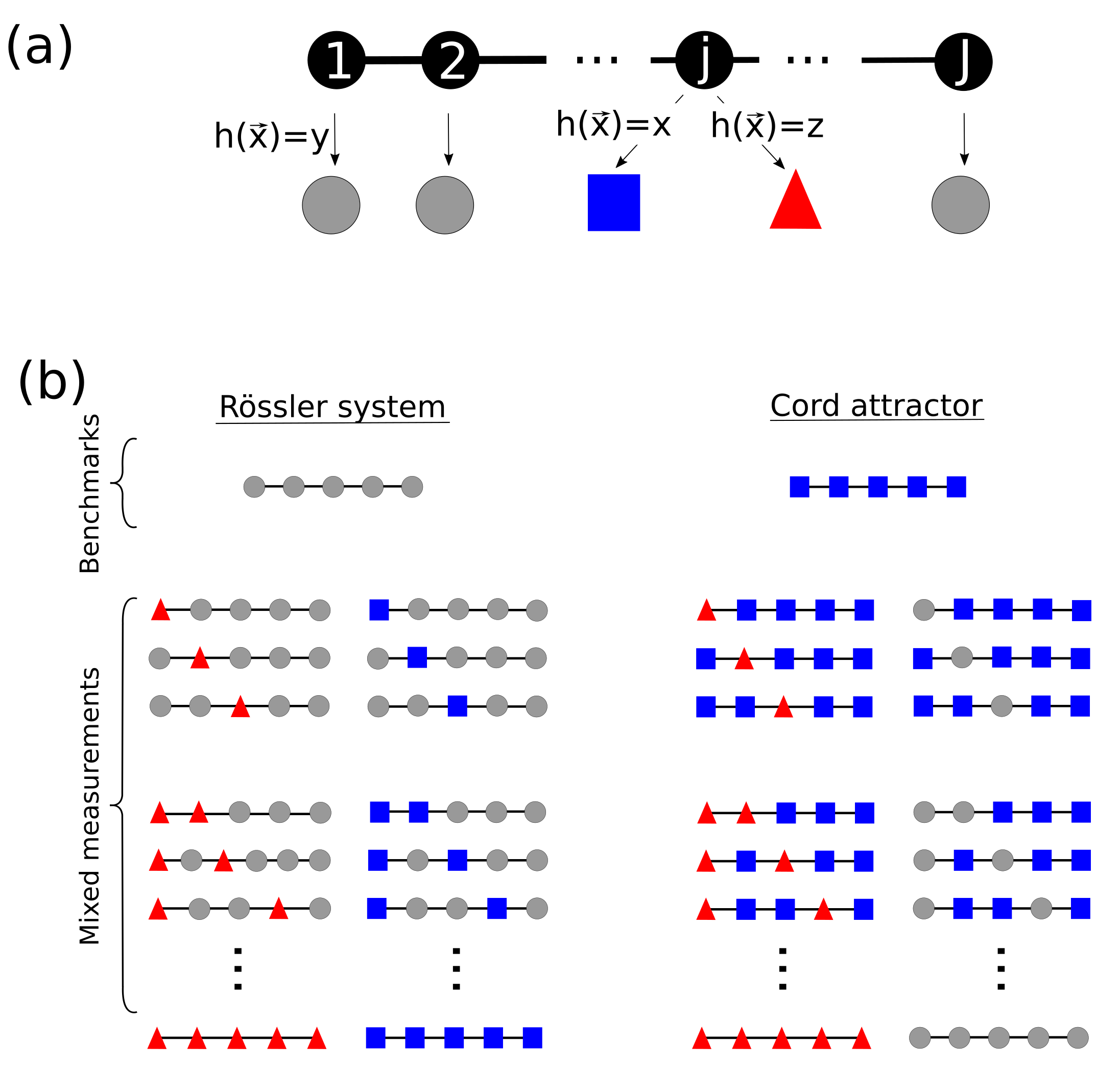}
\caption{\label{fig.scheme} 
Schematic representation of the numerical setup. (a) A chain of $J$ $3$-dimensional oscillators  observed through the measurement functions $h_j(\mb{x})$, which can either return $y_j$ (gray disk), $x_j$ (blue square) or $z_j$ (red triangle). (b) For the benchmark of each network $h_i=h_j,~\forall i,\,j$, and the recorded variable is the one that provides 
best observability of the dynamics. For the cases of
mixed measurements the experimental design explores both the position and number of lower-ranked variables in terms of observability.
Chain networks of $J=5$  R\"ossler and Cord systems are considered.}
\end{figure}

For the sake of presentation, in the following we consider the observability ranking of variables 
for the R\"ossler system \cite{Letellier2005}: $y\rhd x\rhd z$, i.e. $y$ is the best observable, 
followed rather closely by $x$ and, finally, by $z$ which is poor variable 
from an observability point of view. 
Suppose that, for $l<J$ oscillators, only $x$ or $z$ are available. This results in a {\it mixed} measurement set, that will be indicated as $H_{x^l}=\{y_j\}_{j\in\Psi}\cup\{x_j\}_{j\in\Upsilon}$ when mixing  $y$ and $x$ variables, or $H_{z^l}=\{y_j\}_{j\in\Psi}\cup\{z_j\}_{j\in\Upsilon}$ when mixing $y$ and $z$. Hence, the set $\Psi$ (size $J-l$) includes those oscillators from which the $h(\mb{x})=s=y$ variable was measured, while $\Upsilon$ (size $l$) corresponds to the cases for $s\neq y$, so $\Psi\cap \Upsilon=\{\}$. 

Here, we will not consider mixed measurement sets with  all the  state variables (e.g., $H_{x^{l_1},z^{l_2}}$). As shown latter, the results with two poor variables in terms of observability, or too good ones, are almost equivalent. Explicitly, in the scenario of the previous example, the measurement set $\{y_1, x_2, x_3, z_4\}$ would have the equivalent impact on the svM-SSA as the sets $\{y_1, y_2, y_3, z_4\}$  or $\{x_1, x_2, x_3, z_4\}$ because $y$ and $x$ provide good observability of the dynamics.

Three benchmark scenarios will be considered:

\begin{itemize}
\item[(a)] a chain of 5 R\"ossler oscillators with chaotic phase coherent dynamics;
\item[(b)] as for (a) but with funnel (phase incoherent) dynamics; 
\item[(c)] a chain of 5 cord attractors.
\end{itemize}

For the cord attractor the observability order is  $x\rhd y\approx z$ \cite{Aguirre2011}, which
in a sense, is complementary to that of the R\"ossler oscillator because it has two variables that
convey poor observabillity of the dynamics. The mixed measurement sets that will be investigated are shown schematically in Figure~\ref{fig.scheme}(b). It is worth noticing that in all numerical experiments the coupling variable is always the same
for a given network. More details are given in Sec.~\ref{sec.equations}.

\subsection{Multivariate singular spectrum analysis}
\label{sec.mssa}

The varimax approach for the multivariate singular spectrum analysis (M-SSA) for phase synchronization phenomena was originally proposed by Groth and Ghill \cite{Groth2011}. The analysis from a single variable has been investigated in \cite{Portes2016b}. Let $\{s_j(k)\}_{k=1}^N$ be the measured time series of each $j=1, ..., J$ oscillator at time $k$. First the individual trajectory matrices ${\bf X}_j$ should be formed, by embedding each $\{s_j(k)\}$  in an  $m$-dimensional space with lag $1$ \cite{Takens1981,Broomhead1986} ($m$ is also called window width), and then concatenated to form the augmented trajectory matrix ${\bf X}=[{\bf X}_1, ..., {\bf X}_J] \in \bbR^{N-m+1, Jm}$. 

A ``skeleton" of the structure encoded in the time series is extracted by performing a singular value decomposition (SVD) of ${\bf X}$ or, equivalently, the eigendecomposition of the covariance matrix ${\bf C}={\bf X}^T{\bf X}/(N-m+1)$ as ${\bf C}={\bf \Lambda}{\bf E}$ (used in the present work). Due to a known mixture of the eigenvectors related to the individual subsystems, Groth and Ghill introduced a specialized varimax rotation on the first $S$ eigenvectors ${\bf E}_S^*={\bf E}_S{\bf T}$, and the computation of the {\it modified} variances (new singular values) $\{\lambda_k^*\}_{k=1}^S\equiv {\rm diag}({\bf \Lambda}_S^*)$ as ${\bf \Lambda}_S^*={\bf T}^T{\bf \Lambda}_S{\bf T}$. Recently, a class of orthogonal structured rotations was proposed \cite{Portes2016a}, of which the ``structured" varimax \cite{Groth2011} is a special case. Then, the procedure is here referred to as the structured-varimax M-SSA (svM-SSA). 
 
\begin{figure}\centering
\includegraphics[width=0.95\columnwidth]{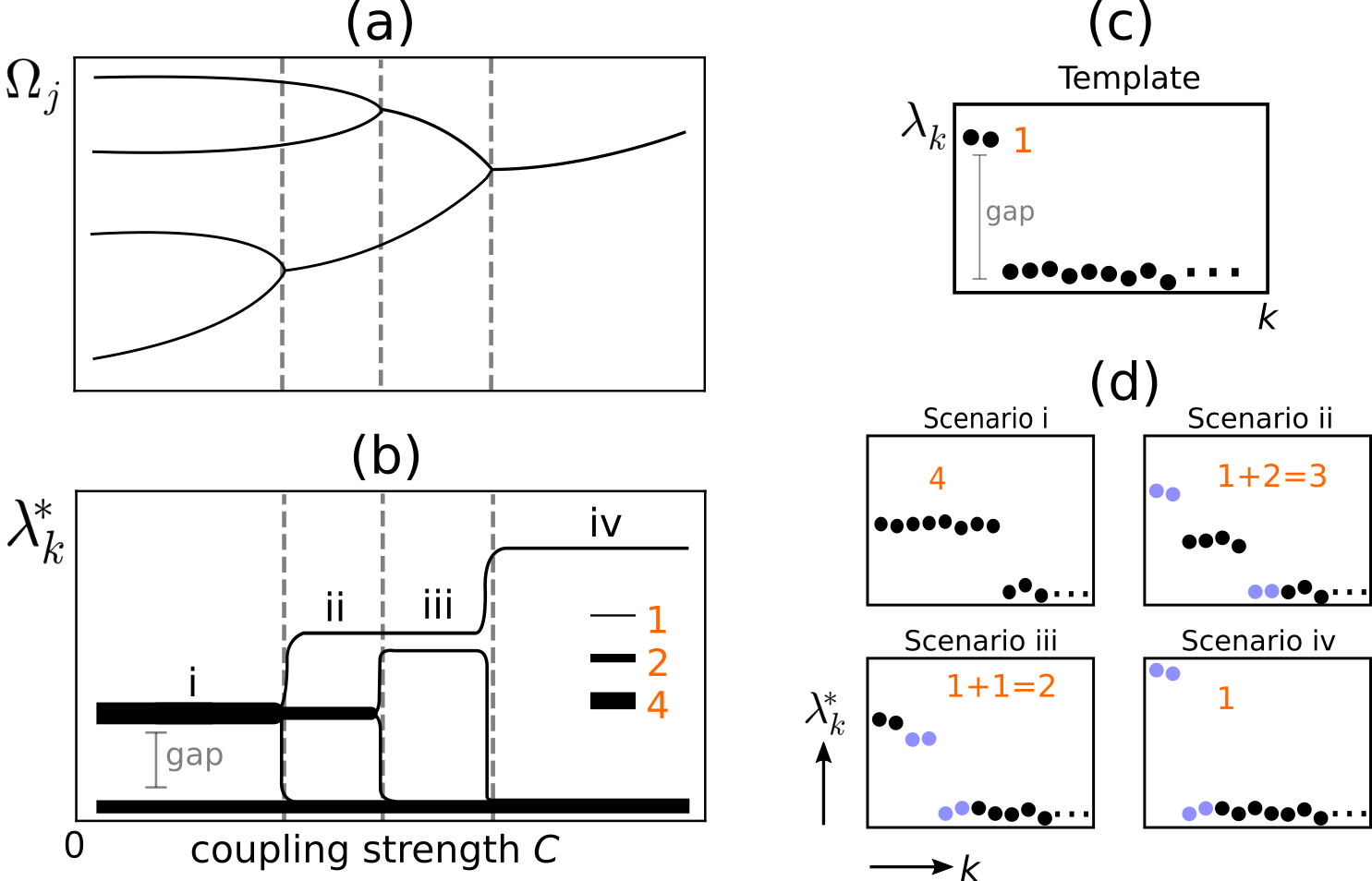}
\caption{\label{fig.mssa.tutorial}Schematic representation of phase synchronization analysis with the svM-SSA for $J=4$ {\it idealized} detuned phase-coherent oscillators. (a) Wihout coupling each oscillator
has its own frequency. With increasing $C$ frequency locking happes at values of the coupling indicated
by vertical dashed lines. (b) The corresponding signatures in the svM-SSA spectrum, where each detected oscillatory mode  is represented by a $\lambda^*_k$ pair [line thickness, and Arabic numerals in (b-d), indicate the number of pairs above the ``noise" floor]. (c) The specific M-SSA signature of {\it a single oscillator}\, (the Template) shows a single $\lambda^*_k$ pair. (d): in scenario (i) there are four unsynchronized oscillators -- compare to the template of a single one in (c); in scenario (ii) two
such oscillators form a PS-cluster that oscillate coherently and therefore is represented by a single
$\lambda^*_k$ pair; in scenario (iii) the  two oscillators that were uncoherent in scenario (ii) become
phase-synchronized, hence there are two clusters of two oscillators; finally, in scenario (iv) the four oscillators become phase-synchronized.}
\end{figure}

The singular values ${\bf \Lambda}_S^*$ provide information of the underlying structure in the data: a single high value is related to a trend; pairs of nearly equal values reflect oscillatory modes; near zero values are associated with noncoherent oscillations and will be referred to as the noise floor. The analysis of phase synchronization, {\it without} having to explicitly estimate phases, is then performed as schematically illustrated in Fig.~\ref{fig.mssa.tutorial} for $J=4$ idealized coherent oscillators.

\subsection{Network Models}
\label{sec.equations}


A chain of $J$ detuned R\"osslers oscillators, diffusively coupled through the $y$ variable 
can be represented as \cite{Osipov1997}:

\begin{eqnarray}
\label{eq.coupRossler}
\left\{ 
\begin{array}{l}
\dot x_j=- \omega_j y_j-z_j,\\
\dot y_j=\omega_j x_j+a y_j+C (y_{j+1}-2y_j+y_{j-1}),\\
\dot z_j=0.1+z_j(x_j-8.5),
\end{array} \right.
\end{eqnarray}
where $C$ is the coupling strength, $\omega_j \!=\! \omega_1 \!+\! \Delta\omega(j\!-\!1)$ are the natural (intrinsic) frequencies with $\omega_1=1$, $\Delta\omega=0.02$ and the index $j=1, ..., J$ being the position in the chain, with free boundary conditions $y_0=y_1$ and $y_{J+1}=y_J$. In this case the vector of coupling variables is $\mb{v}=[y_{j-1} ~y_{j+1}]^T$. 
The observability rank is $y\rhd x\rhd z$ with $y$ and $x$ being  good variables to reconstruct the dynamics, and $z$ a poor one \cite{Letellier2005}.

The impact of a mixed measurement in PS characterization was investigated in two different chaotic regimes, regarding phase coherence. The first one is the phase-coherent chaotic spiral regime ($a=0.15$), for which the trajectory projection onto the $xy$-plane has a well defined center of rotation, and a single  dominant time scale as reveled by both the power spectrum density (PSD) and the svM-SSA template  (see Fig. 2 (a-c) in \cite{Portes2016b}).
The second one is the chaotic funnel regime ($a=0.28$) with no phase coherence, no center of rotation, and exhibiting several time scales as seen in its PSD and svM-SSA template (see Fig. 2(d-f) in~ \cite{Portes2016b}).

System (\ref{eq.coupRossler}) was integrated with time step $t_{\rm int}=0.01$ t.u. (time units), for a total simulation time $t_{\rm sim}=12\times 10^3$ t.u. The initial transient $t_{\rm trans}=2\times 10^3$ t.u. was removed, and the time series of $x$, $y$ and $z$ of each oscillator were sampled with sampling time $t_{\rm s}=0.4$ t.u.
 This was done for $200$ linearly spaced values of the coupling strength $C\in[0,0.1]$ for the spiral chaotic regime, and $C\in[0,0.62]$ for the funnel one (since a larger coupling strength was necessary to synchronize the intrinsic funnel oscillators).


A chain of $J$ diffusively $x$-coupled chaotic cord systems can be represented as:
\begin{eqnarray}\label{eq.cord}
\left\{ 
\begin{array}{l}
\dot x_j= -y_j -z_j -ax_j +aF +C (x_{j+1}-2x_j+x_{j-1}),\\
\dot y_j= x_j y_j -b_j x_j z_j -y_j,\\
\dot z_j= b_j x_j y_j +x_j z_j -z_j, 
\end{array} \right.
\end{eqnarray}
with $(a,F,G)=(0.258, 8.0, 1.0)$ and free boundary conditions $x_0=x_1$ and $x_{J+1}=x_J$. The detuning of the intrinsic temporal scales was done by setting $b_j=0.4033+(j-1)\Delta b$ with $j=1,...,J$ and $\Delta b=0.016$. Here the vector of coupling variables is $\mb{v}=[x_{j-1} ~x_{j+1}]^T$. 
The simulations were performed with $(t_{\rm int},t_{\rm s},t_{\rm sim},t_{\rm trans})=(0.01, 0.02,4000,2000)$, for $200$ values of an increasing (and logarithmically spaced)  coupling strength $C\in[0.001,0.1]$.

The cord oscillator is a more challenging case because it possess only one variable that provides good observability of the system dynamics, with the observability rank $x\rhd y\approx z$ (being $y$ and $z$ poor variables) \cite{Aguirre2011}. The trajectory projection onto the $xz$ and $yz$ planes reveals that  the information about the ``cord", linking both sides of the attractor, is completely lost without information from $x$ (see Fig. 7 in~ \cite{Portes2016b}).
%

\section{Results}
\label{sec.resulsts}

\subsection{R\"ossler system: phase-coherent regime}
\label{sec.rossler}

\begin{figure} 
		\centering
		\includegraphics[width=0.95\columnwidth]{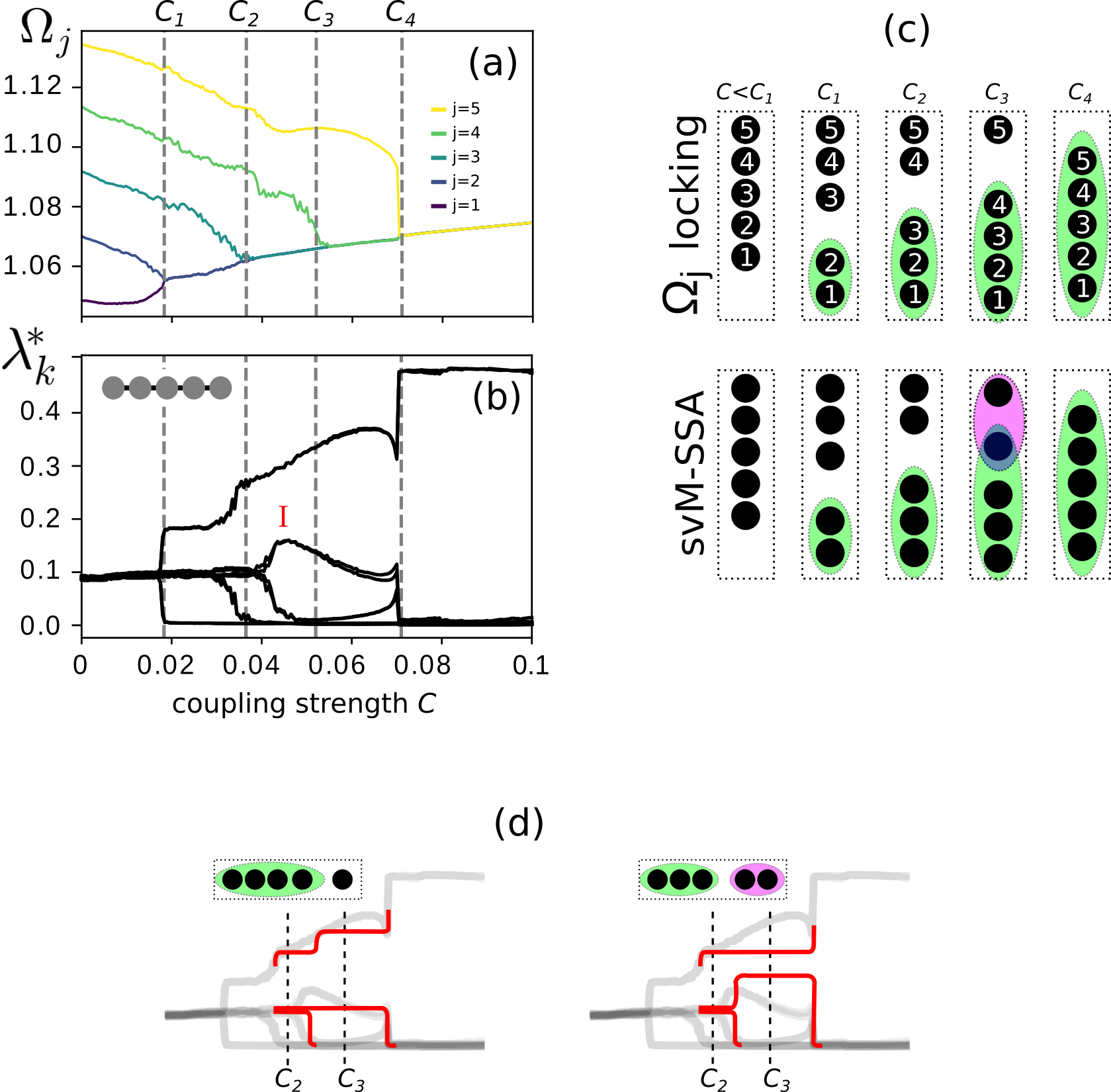}
		\caption{\label{fig.rossler-coherent-5y}Synchronization in a chain of $J=5$ $y$-coupled R\"ossler oscillators in the {\it spiral} chaotic regime. The (a) mean observed frequencies, estimated as $\Omega=\langle \dot{\phi_j}\rangle_t$ with $\phi_j=\arctan(y_j/x_j)$, and the (b) singular values from the svM-SSA of the $y$ time series -- vertical gray lines mark the onsets of frequency locking, as suggested by $\Omega$. Both analyses show a cascade of PS cluster formation, schematically represented in (c,\,d). But while the (a) mean frequency analysis is the only able to identify the oscillators, the (b) svM-SSA is able to identify an intermittent PS range between $C_2$ and $C_3$ [red ``I' in (b)] -- further analysis of the estimated phases $\phi_j(t)$ (not shown), reveals that oscillator $j=4$ is intermittently phase synchronized with the PS cluster $j=\{1,2,3\}$ and with oscillator $j=5$. For the sake of illustration, $\lambda_k^*$ spectrum signature for alternative non-intermittent behaviors (red lines) are shown in (d).}
\end{figure}

Figure~\ref{fig.rossler-coherent-5y} shows the synchronization analysis of the benchmark measurement set (i.e., all $y$). The mean observed frequencies $\Omega_j$ are presented, along with the svM-SSA $\lambda_k^*$ spectrum, as an auxiliary metric to support the discussion. 
Both analyses suggest that as the coupling strength $C$ is increased, phase-synchronized clusters
are gradually formed up to the point $C\approx 0.07$ where all oscillators finally become phase-synchronized.
The onset of frequency locking (vertical dashed lines) is in agreement with the onsets of PS suggested by the $\lambda_k^*$ spectrum at $C=\{C_1,C_2, C_4\}$, which are identified by the increasing value of a $\lambda_k^*$ pair with a simultaneous drop of other one to the noise floor.

In the range between $C_2$ and $C_3$ (marked by the red ``I"), the  simultaneous increase of two $\lambda_k^*$ pairs, along with the dropping of a third one is seen. This suggests an intermittent synchrony behavior, as pictorially illustrated in Fig.~\ref{fig.rossler-coherent-5y}(c)  (which was confirmed by a straightforward analysis of phase difference, not shown).  For $C\approx C_3$, there is, on the one hand, a PS cluster of three oscillators
and, on the other, a single oscillator; and the remaining oscillator intermittently synchronizes with each of these. 
The situation in which it synchronizes with the cluster is indicated by the green set in  
Fig.~\ref{fig.rossler-coherent-5y}(d),  and when it synchronizes with the single oscillator 
is indicated by the purple set in Fig.~\ref{fig.rossler-coherent-5y}(d).
This intermittency is not captured by the mean frequency locking analysis. Fig.~\ref{fig.rossler-coherent-5y}(d) shows in red lines the results that would be expected if no intermittent behavior took place.

%

\begin{figure}
  \centering
  \includegraphics[width=0.95\columnwidth]{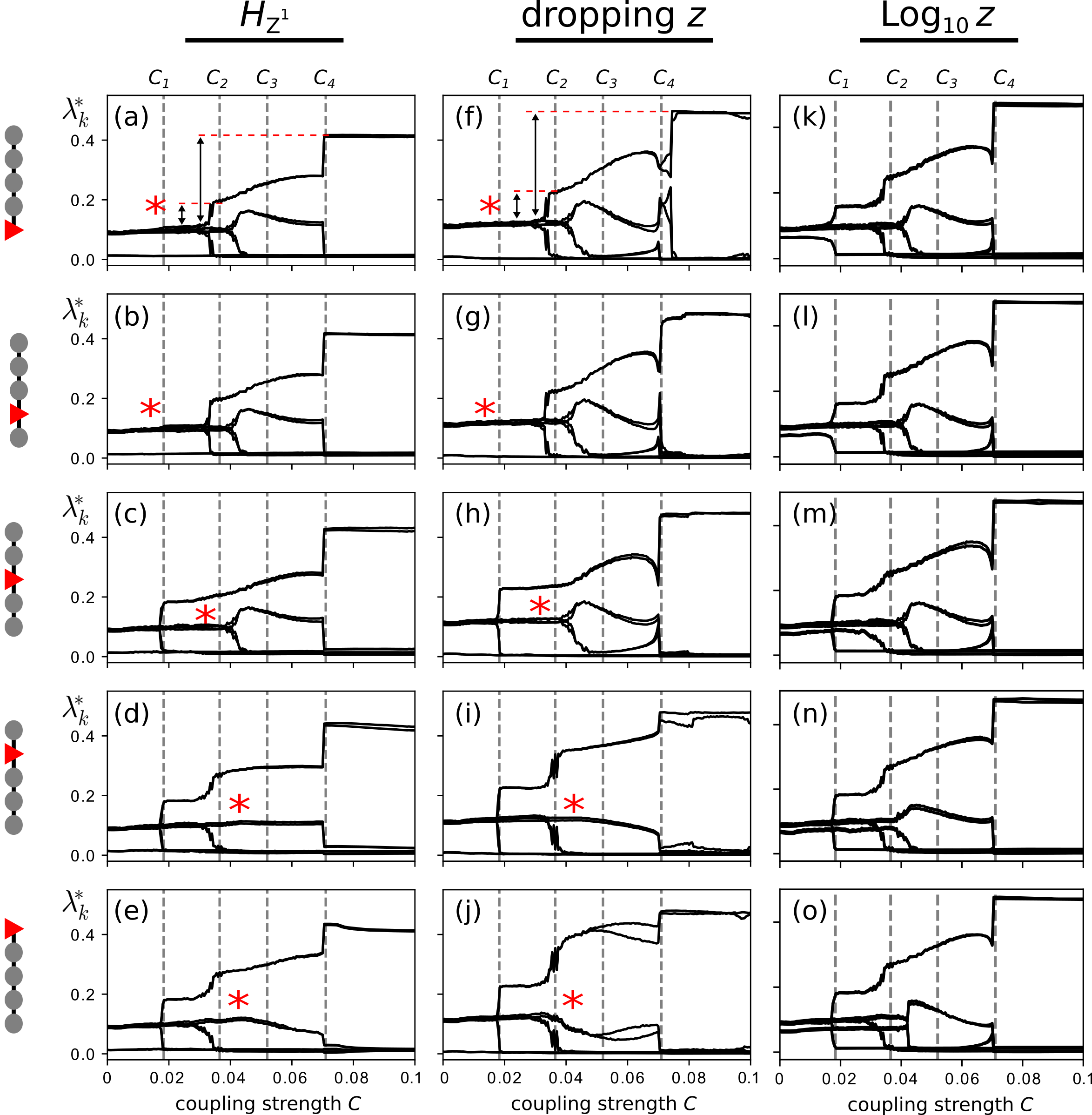}
  \caption{\label{fig.rossler-coherent-1z}(a-e) Impact of a single $z$ measurement of an oscillator -- the other four being measurements of the $y$ variable of the other four oscillators -- in the synchronization analysis of a chain of $J=5$ R\"ossler oscillators in the spiral chaotic regime. The vertical gray lines are as in Fig.~\ref{fig.rossler-coherent-5y}. The red star indicates missing features. Roughly, the oscillator observed through $z$ becomes ``invisible" to the svM-SSA. (f-j) A similar situation is found when the measurement of $z$ is removed from the analysis, but with some improvement in resolution enhancement [see vertical arrows, only showed in (a) and (f)]. (k-o) The use of $\log_{10} z$, instead of $z$, fully recovers the overall features of the benchmark (see Fig.~\ref{fig.rossler-coherent-5y}b).}
 \end{figure}

Those aforementioned landmark events in the PS dynamics of the network are not fully captured by using a mixed measurement set that includes the $z$ variable. First, consider the results for the five possible $H_{z^1}$ scenarios, Fig.~\ref{fig.rossler-coherent-1z}(a-e). The oscillator measured through $z$ becomes ``invisible", in the sense that the events that happen at the location indicated by red stars in the $\lambda_k^*$ spectrum disappeared [see the benchmark in Fig.~\ref{fig.rossler-coherent-5y}(b)]. Indeed, these results are similar to the ones obtained by explicitly {\it removing} the respective $z$-measured oscillator from the analysis, as shown in Fig.~\ref{fig.rossler-coherent-1z}(f-j).

Another relevant feature is that, rather surprisingly, the results provided by this latter approach are closer to the benchmark ones. Notice, for instance, the sudden drop of the singular values just before the onset of global PS ($C=C_4$), which is absent in all the $H_{z^1}$ scenarios, but is present when one ``ignores" the oscillators $j=1, 2$ or $3$, Fig.~\ref{fig.rossler-coherent-1z}(f-h). Also, the resolution is somewhat improved as revealed by a larger gap between the leading eigenvalues and the noise floor [e.g., compare the vertical arrows in Fig.~\ref{fig.rossler-coherent-1z}(a) and (f)].
Hence, in the present example, the results suggest that some features become more evident when a poor observable 
is completely left out of the analysis. 
 
It is known that $\log_{10} z$ provides good observability from the system dynamics \cite{Ibanez2006,Aguirre2011}, as oposed to the $z$ variable. In order to double-check the hypothesis concerning observability, the M-SSA was performed using $\log_{10} z$ instead of $z$ for the indicated oscillators.  The plots in Fig.~\ref{fig.rossler-coherent-1z}(k-o) show the results obtained. No PS landmark is missing now, and the overall quality of the spectrum is similar to the benchmark one, Fig.~\ref{fig.rossler-coherent-5y}(b).

\begin{figure}
		\centering
		\includegraphics[width=0.5\columnwidth]{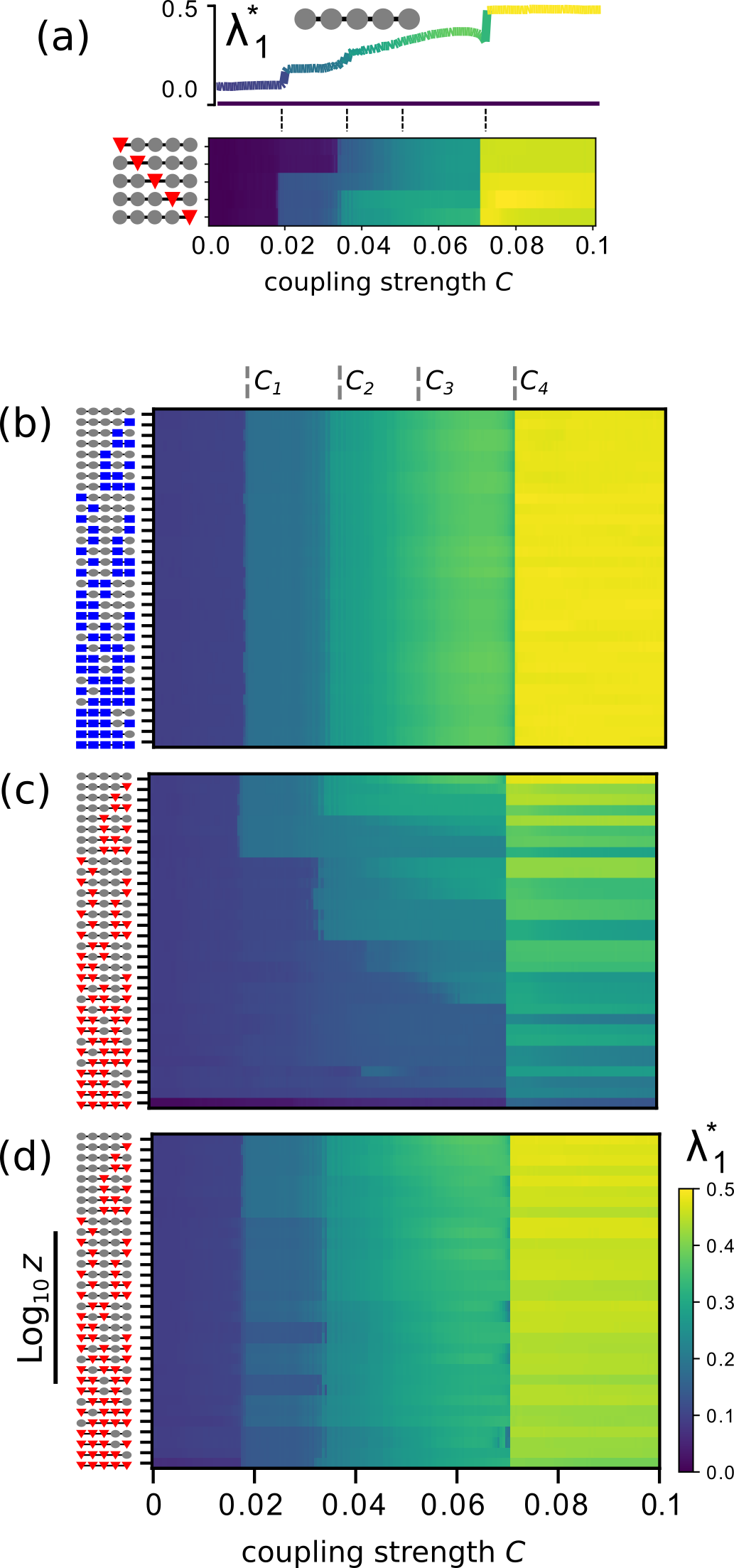}
		\caption{\label{fig.rossler-coherent-image-plot} Impact of mixed measurements for a chain of $J=5$ R\"ossler oscillators in the {\it spiral} chaotic regime. (a) The leading singular value $\lambda_1^*$ provides a compact representation for the mixed scenarios, as exemplified with $H_{z^1}$ [see Figs.~\ref{fig.rossler-coherent-1z}(a-e)]. The color-map is provided by $\lambda_1^*$ of the ``benchmark" scenario (only $y$) which can be seen in the first row
of the color-maps. The results with (b) $x-y$ are almost identical to those using only $y$. The mixing with (c) $y-z$ shows the deterioration of its discriminatory power, e.g. the transition between $C_1$ and $C_2$ is not always detected. (d) Results using $\log_{10} z$, instead of $z$, are similar to those with $y$ and $x$.}
\end{figure}

In order to express the results for all the $32$ possible measurement sets, in a concise way, it is worth noticing that $\lambda_1^*$ (the highest singular value) provides a pertinent amount of information about the PS behavior, as shown in Fig.~\ref{fig.rossler-coherent-1z}(f). The $\lambda_1^*$ from the benchmark is used to set the color scale (upper panel), and the matrix plot of the $\lambda_1^*$ from scenarios $H_{z^1}$ (lower panel) clearly agrees with the earlier discussion of the full svM-SSA  plots [Figs.~\ref{fig.rossler-coherent-1z}(a-e)].
In view of this, Figs.~\ref{fig.rossler-coherent-image-plot}(a,\,b) show the results for all combinations of mixed measurements with the $z$ and $x$ variables, respectively. When the majority of oscillators are measured through $z$ (e.g. in $H_{z^5}$ and $H_{z^4}$), only the transition to global PS (one cluster) is detected (with a low contrast). 
The results with $H_{x^l}$ are equivalent to that of the benchmark. In the view of \cite{Portes2016b}, this is expected, since both $x$ and $y$ variables provide good observability of the system dynamics, and being equivalent for the svM-SSA in a phase coherent dynamics. That is not the case for the noncoherent phase funnel regime, as shown it the next section.

It is worth noting that the $z$-measured oscillators positions in the chain, and the number of them, are not the relevant, determining aspect to consider. For example, the first emergent clustering at $C=C_1$ is detected by using any mixed measurement set, provided the oscillators that become part of the cluster ($j=1,\,2$) are not being measured through the poor variable $z$ (as seen in Fig.~\ref{fig.rossler-coherent-image-plot}(b).

\subsection{R\"ossler system: funnel regime}
\label{sec.rossler-funnel}
As discussed in Sec.~\ref{sec.equations}, the R\"ossler oscillator dynamics in the chaotic funnel regime presents several time scales and, accordingly, its characteristic fingerprint is not a single  $\lambda_k^*$ pair anymore (as in the chaotic spiral regime), but several pairs followed by a slowly decreasing tail. In practice, however, synchronization analysis could be carried out by considering just the two higher $\lambda_k^*$ pairs, as they have an appropriate vertical distance from each other and from the noise floor (see Fig. 2(f) in~ \cite{Portes2016b}).
 
\begin{figure}
		\centering
		\includegraphics[width=0.5\columnwidth]{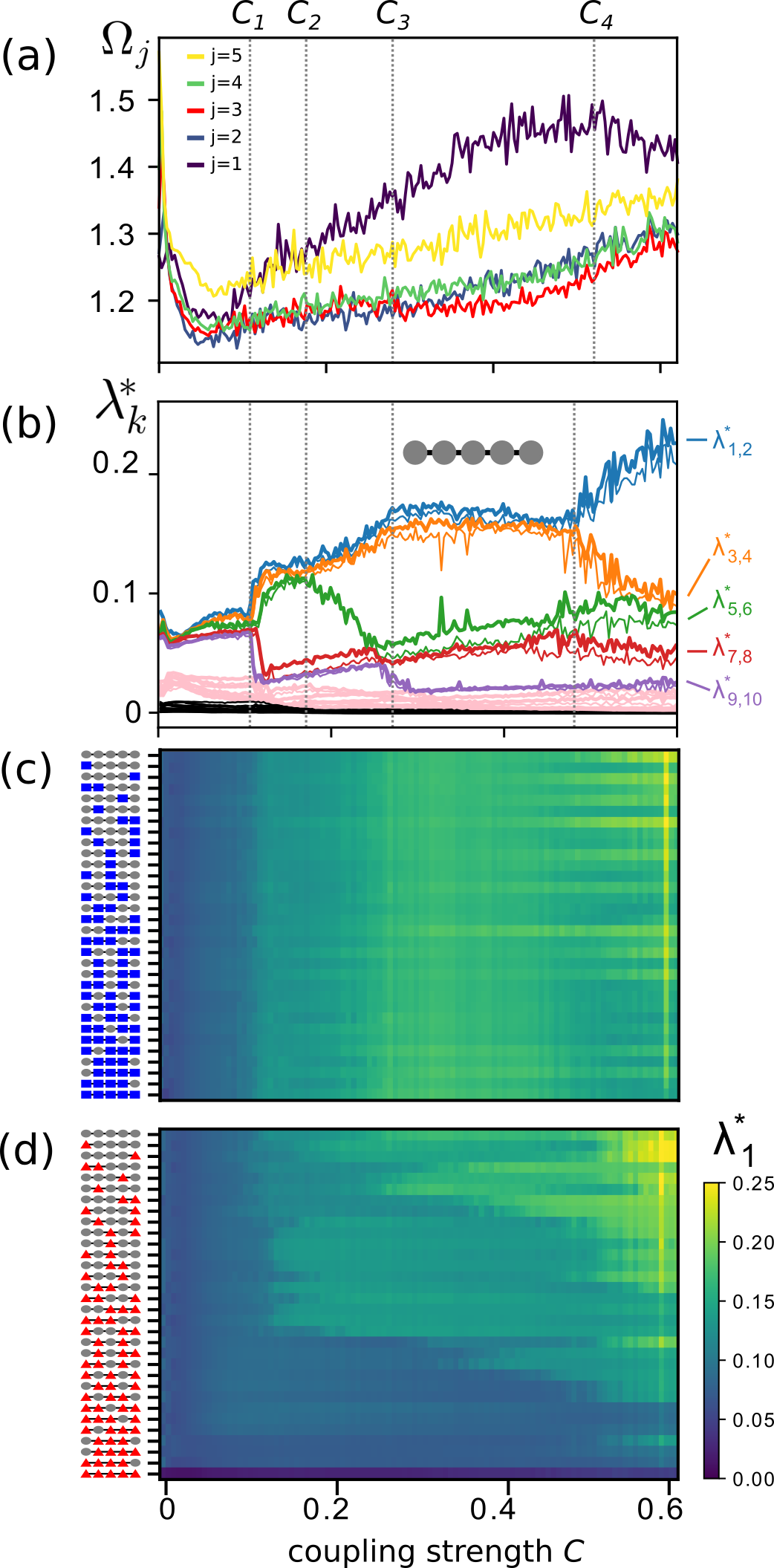}
		\caption{\label{fig.rossler-funnel-plot}Phase synchronization in a chain of $J=5$ R\"ossler oscillators in the {\it funnel} chaotic regime. The (a) mean observed frequencies, estimated as $\Omega=\langle \dot{\phi}_j\rangle_t$ with $\phi_j=\arctan(\dot{y}_j/\dot{x}_j)$, and the  (b) singular values from the svM-SSA of the $y$ time series -- vertical lines indicate four arbitrary salient PS landmarks to guide visual inspection. The leading singular value $\lambda_1^*$ for (c) $x-y$ mixing are similar to those that only use $y$, whereas the results for (d) $y-z$ shows the deterioration of its discriminatory power.}
\end{figure}

The results for the benchmark measurement set are shown in Figs.~\ref{fig.rossler-funnel-plot}(a,\,b).  A clear tendency to PS is suggested by the $\lambda_k^*$ spectrum, and four landmark PS-related events (dashed vertical lines) were selected from it -- note that the mean frequency locking analysis, shown here for the sake of completeness, is far less clear in the funnel regime. The onset of PS is near $C=C_1$: from the ten leading $\lambda_k^*$ {\it pairs} at $C<C_1$, six become larger while the other four drop to the noise floor. The same occurs for the pairs $\lambda_{11,..., 20}^*$, related to the second oscillatory mode of each oscillator (but their low values make the visualization difficult). Similar events occur at $C=\{C_2, C_3,C_4\}$. 

For the funnel regime, the mixed measurements with $y$ and $x$ do not yield the same results as when only $y$ is used. This can be better appreciated by contrast to the results for the spiral regime shown in Fig.\ref{fig.rossler-coherent-image-plot}(b) where all the rows are basically the same regardless of the combination of $y$ and $x$ measurements. In the case shown
in Fig.~\ref{fig.rossler-funnel-plot}(c) not all the rows are equivalent to the first one (the benchmark).  This seems to be a consequence of the slightly higher observability provided by $y$ as compared to $x$, which was recently found to enhance the svM-SSA in the context of no-coherent phase dynamics~\cite{Portes2016b}.
Finally, the use of $z$ hinders the PS clustering analysis by means of the M-SSA, as seen in Fig.~\ref{fig.rossler-funnel-plot}(d).

%

\subsection{Cord oscillator}
\label{sec.cord}

\begin{figure}
  \centering
  \includegraphics[width=0.5\columnwidth]{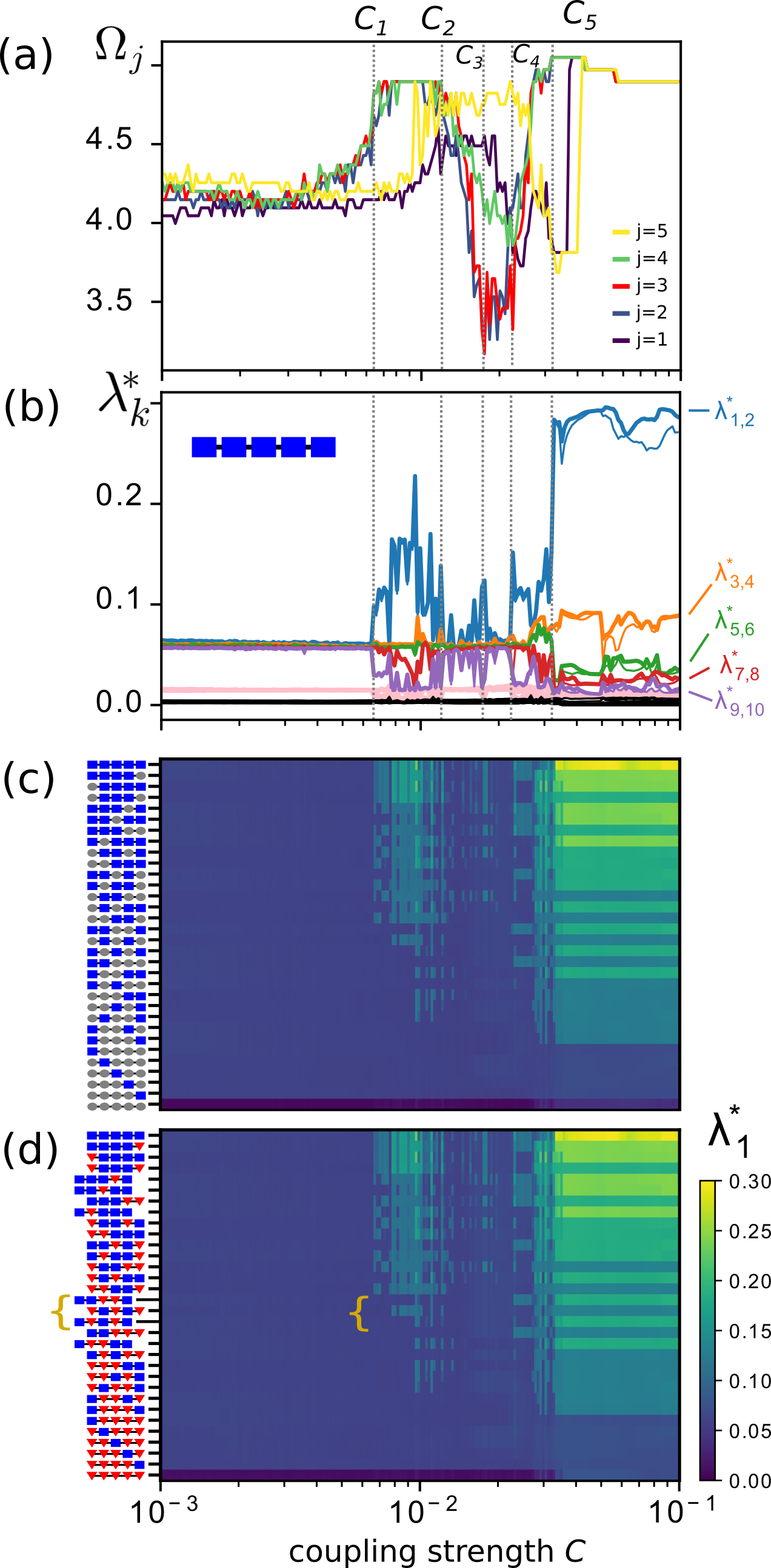}
  \caption{\label{fig.cord-image-plot}Phase synchronization in a chain of $J=5$ chaotic cord oscillators. (a)~Mean observed frequencies computed through  phase estimates taken with the Poincar\'e section $\mathcal{P}=\{(y,z)\in\mathbb{R}^2|x=0,\dot{x}>0\}$. (b)~Singular values from the svM-SSA, computed with the $x$ time series (the benchmark).
The analysis through the leading $\lambda_1^*$ shows that mixed measurements with (c)~$z$ and (d)~$y$ have similar detrimental effects.}
\end{figure}

The  $x$ variable is the only one to provide good observability of the cord attractor dynamics. The respective synchronization analysis, used as the benchmark, along with the mean observed frequency analysis is shown in Figs.~\ref{fig.cord-image-plot}(a,\,b). 
The complex synchronization dynamics hinders the ``standard" straightforward frequency locking analysis. However, some landmark events -- marked by dashed vertical lines at $C=\{C_1, ..., C_5\}$ -- are clearly identified by the svM-SSA, as discussed next.

First, the onset of a PS cluster occurs at $C\approx C_1$, which dies out at $C\approx C_2$. It is formed through a PS clustering cascade: (i) notice in Fig.~\ref{fig.cord-image-plot}(a) the successive frequency locking of oscillators $j=3,4$ and later of $j=2$; (ii) accordingly, the svM-SSA in Fig.~\ref{fig.cord-image-plot}(b) shows two $\lambda^*$ pairs successively dropping to the noise floor (arbitrarily labeled as $\lambda^*_{7,8}$ and $\lambda^*_{9,10}$) -- other specific events can be seen, as oscillator $j=5$ intermittently entering the cluster between $C_1$ and $C_2$, and oscillator $j=2$ intermittently leaving the cluster. These features, identified through the mean frequency analysis, have their respective fingerprints in the svM-SSA spectrum as two $\lambda^*$ pairs that intermittently go to, and return from, the noise floor. However, in the latter plot $\lambda^*$ pairs cannot be identified to specific oscillators.

Second, increasing the coupling strength further,  two salient peaks in the $\lambda^*$ spectrum at $C_3$ and $C_4$ are evidence of PS onsets. Finally, global PS emerges at $C_5$, as seen by the single high $\lambda^*_{1,2}$ pair (both this pair and the second leading one, $\lambda^*_{3,4}$ are the specific signature of a single cord attractor, see Fig.~7(c) in \cite{Portes2016b}). In fact, the ``quality" of this global PS varies at higher values of the coupling, as suggested by the fluctuations in the $\lambda^*$ spectrum. The mean frequency analysis detects this event much latter, due to the occurrence of phase slips (not shown).
 
Observability analysis  \cite{Aguirre2011,Letellier2005} provides similar low observability levels for both $y$ and $z$ variables. In view of this, the detrimental effect due to mixed measurements scenarios $H_{y^n}$ and $H_{z^l}$ are also expected to be equivalent. This prediction is fully supported by the almost identical results shown in Fig.~\ref{fig.cord-image-plot}(c,\,d)  (e.g., notice the small difference for $H_{y^5}$ and $H_{z^5}$  at $C\approx C_5$). Specifically, the aforementioned PS events become undetectable when the {\it relevant} oscillators are measured through $y$ or $z$, which
convey poor observability.

\begin{SCfigure*}
  \centering
  \caption{\label{fig.cord-no-z}Phase synchronization in a chain of $J=5$ chaotic cord oscillators for selected cases corresponding to the left-shifted symbolic sets in Figs.~\ref{fig.cord-image-plot}(c).  Dropping the $z$-measured oscillator from the analysis provides results that are similar to those computed with mixed measurements with (e)~one and (f)~two $z$ variables, but a enhanced resolution is achieved (vertical arrows).}
  \includegraphics[width=0.7\textwidth]%
    {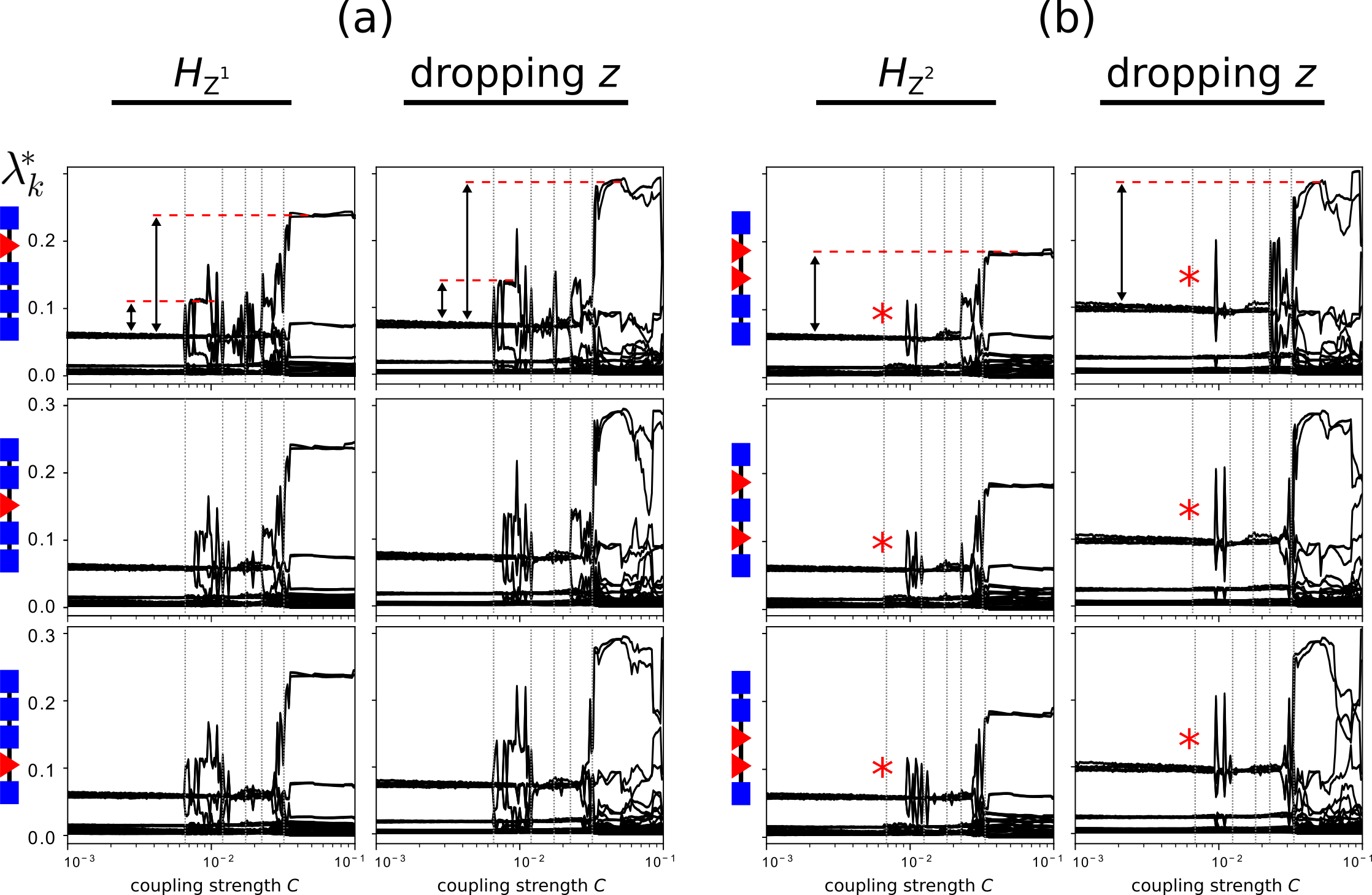}
\end{SCfigure*}

In order to investigate  this effect further, and compare the results with the strategy of explicitly exclude the oscillators measured through poor variables prior the svM-SSA, six mixed measurement sets with one {\it or} two $z$ variables were selected -- which correspond to the scenarios marked by left-shifted symbolic set labels in Fig.~\ref{fig.cord-image-plot}(c).  The corresponding full svM-SSA spectra are shown in Figs.~\ref{fig.cord-no-z}(a,\,b).
 The following features are worth noticing. 
 First, the overall aspect of a given svM-SSA spectrum computed through a mixed measurement set and the corresponding ``$z$-dropped" one is quite similar, but the latter provides a somewhat enhanced resolution: for example, the vertical arrows, corresponding to the gap of the leading $\lambda^*$ between two prominent PS clustering scenarios and the non-synchronized one, are larger when $z$ is dropped [for the sake of clearness, the arrows are shown only in the first row of  Figs.~\ref{fig.cord-no-z}(a,\,b), but the enhanced resolution can be seen in all the other plots].
 Second, since the first PS cluster is formed by the interplay of {\it three} oscillators, its signature in the $\lambda^*$ spectrum disappears (red stars) only when at least {\it two}\, oscillators of this cluster are measured through $z$ or simply not measured.

\section{Discussion}

The simulation results highlight two points. First, an enhanced resolution (i.e., a larger gap between the noise floor and the leading eigenvalues) in the detection of PS landmark events was possible by dropping out the poor variables from a mixed measurement set. In other words, it {\it could}\, be a better strategy to not measure a given set of nodes instead of doing so through poor variables. 
 Second, the use of a set of mixed variables affects in {\it different} ways the detection of (i) PS clustering and (ii) global PS, a feature that could be relevant to guide a given experimental design or measurement strategy in real-world applications.

\begin{figure}
  \centering
  \includegraphics[width=0.8\columnwidth]{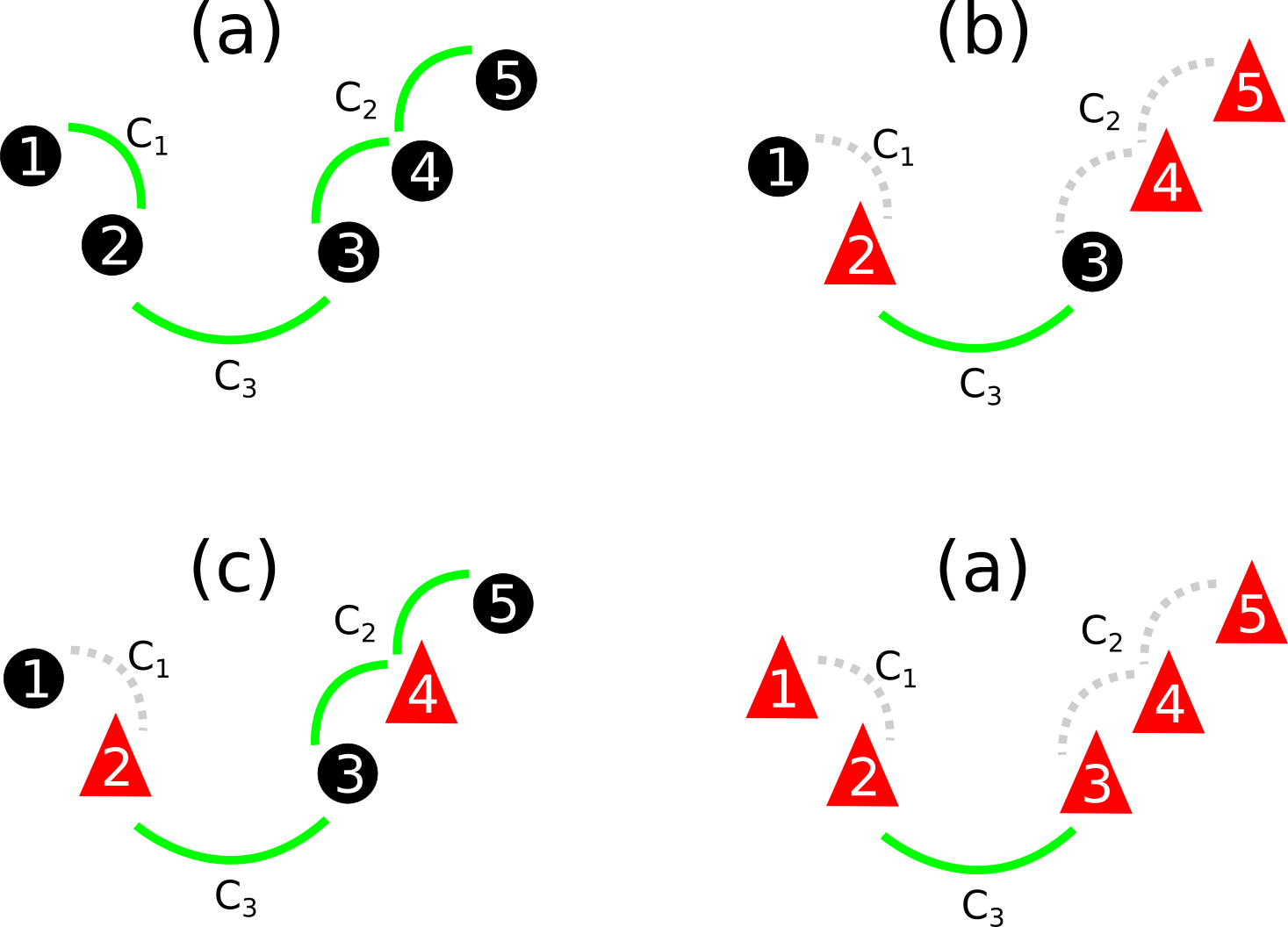}
  \caption{\label{fig.discussion}Schematic representation of a generic $5$-nodes network, passing through a PS cascade $C_1$, $C_2$ until global PS $C_3$. The viable transitions to be detected (green arcs) depend not only on the number of nodes measured through good variables  in terms of observability, but if the PS is limited to a specific cluster or it is achieved for the entire network (i.e., global PS). Black disks and red triangles stand for nodes measured through good and poor variables, respectively.}
 \end{figure}

The formation of a PS cluster can be detected if two conditions are met. First, one must have access to a good variable in terms of observability. Second, at least two nodes in that PS cluster should be measured with that variable, otherwise the synchronization phenomenon is not detectable. To illustrate this point, consider the generic network with five nodes in Fig.~\ref{fig.discussion}(a),
that gradually become PS as indicated schematically by
the green arcs as the coupling is gradually increased $C_1 >C_2>C_3$  until global PS is attained. For instance, if {\it only}\, oscillators at nodes $1$ and $3$ are measured through a good variable [indicated by black discs in Fig.~\ref{fig.discussion}(b)] the event at $C_3$ will be detected, but not events at $C_1$ and $C_2$ (gray dashed arcs). Detection of $C_2$ would be achieved only if nodes $4$ or $5$ [Fig.~\ref{fig.discussion}(c)] were also measured through a good variable.  
 However, an alternative condition applies to the detection of global PS event $C_3$: it can be detected if {\it all} nodes were measured through a poor variable [Fig.~\ref{fig.discussion}(d)].

It is conjectured that the possibility of being able to detect {\it global PS}\, even using variables that convey poor observability is related to the low variance of the data represented by the svM-SSA eigenvectors associated with the poor variables, captured by the $\lambda^*$ spectrum. 
%
In particular, 
global PS implies that a single leading $\lambda^*$ pair accounts for the variance of the entire data set. As a consequence, this pair is much larger than the several  leading pairs of a non-synchronized state scenario, and hence the aforementioned small gap from the noise floor could become noticeable if a {\it sufficiently} large number of oscillators (e.g., the entire network) synchronizes. This is seen in the last six rows of Fig.~\ref{fig.cord-image-plot}(c,\,d), where the transition $C_5$ is not visible with the mixed measurement sets $H_{y^4}$ and $H_{z^4}$, but did become noticeable with $H_{y^5}$ and $H_{z^5}$ (last row).

The described ``all-poor variables" condition is a worse-case scenario. This can be confirmed by noticing that in Fig.~\ref{fig.rossler-coherent-image-plot}(c) the global PS transition $C_4$ is visible by using the $H_{z^4}$ set. In this case, the contributions from the four $z$-measured nodes were sufficient to make noticeable difference in the overall variance represented by the first $\lambda^*$ pair, which  is predominantly determined by the  single $y$-measured node.

\section{Conclusion}
\label{sec.conclusion}

The svM-SSA is a powerful technique to characterize PS \cite{Groth2011} and to provide detailed information of the PS clustering dynamics. The highest discriminatory power is achieved through a single variable approach with the best variable in terms of observability \cite{Portes2016b}. 
However, in practice this specific variable might not be accessible at all nodes of a given network as illustrated recenly by the lack of data from some macroeconomic indicators in the synchronization analysis of world economic activity \cite{Groth2017}. The results in this paper show that although the use of poor variables will not permit the detection of all PS transitions, it could be beneficial in detecting global PS. 

Investigating chain networks of R\"ossler and cord oscillators, this paper has discussed not only some pitfalls that could appear when a mixed set of time series (i.e., a mixed measurement approach) is used, but also the {\it possibility}\, to design more flexible measurement strategies  if one aims at the detection of specific phase synchronization events.

Three features are worth noticing.
 First, the oscillatory modes of a node from which  a poor variable is recorded are not clearly detected and, consequently, landmark PS clustering events could become virtually invisible to the technique.
 Second, the use of the time series of a poor variable could have a detrimental effect on the svM-SSA in the characterization of PS clustering related to the {\it other}\, nodes. This seems to be confirmed by the enhanced resolution obtained in the svM-SSA technique by dropping out the poor variables from the analysis.  In the specific scenario of the chain network of spiral R\"ossler  and cord oscillators, the results suggest that if only variables that convey poor observability of the dynamics
are available at certain nodes, the synchronization analysis with svM-SSA would be preferable without any such measurements. This is more related to numerical features of the svM-SSA than to observability issues.
%
Third, global phase synchronization could be detected in some mixed-measurement scenarios, or even using only poor variables, providing that a sufficiently large number of poor variables is used in order to enhance the discriminatory power of the associated svM-SSA eigenvalue pair.

\section*{ACKNOWLEDGMENTS} 
This work was carried out with CNPq support, Conselho Nacional de Desenvolvimento Cient\'ifico e Tecnol\'ogico - Brazil.






%

\end{document}